# On the detection of relativistic magnetic monopoles by deep underwater and underice neutrino telescopes.

B.K.Lubsandorzhiev

*Institute for Nuclear Research of RAS*

*pr-t 60th Anniversary of October, 7A, 117312 Moscow, Russia.*

Postal address: pr-t 60$^{th}$ Anniversary of October, 7a, 117312 Moscow, Russia; phone: +7-095-1353161; fax: +7-095-1352268; E-mail: lubsand@pcbai10.inr.ruhep.ru

**Abstract**

We present here some reflections and very speculative remarks on the detection of relativistic Magnetic Monopoles by currently operating deep underwater/ice neutrino telescopes.

A possible existence of isolated free magnetic charges and their direct detection have been exhilarating many generations of experimental physicists since as early as the beginning of the 19th century. After P. Dirac's [1,2] introduction of magnetic monopoles (MM) to make Maxwell's equations symmetric and to explain electric charge quantization, a lot of experiments, mainly at accelerators have been carried out to detect MMs, see [3,4] for extensive reviews of the issue.

The advent of the Grand Unification Theory (GUT) of strong and electroweak interactions in the early 70s has inhaled new breath into experimental searches for MMs. At that moment the experimental efforts concentrated mostly on searches for super heavy MMs ensued from GUT theories. The Rubakov-Callan effect [5,6] opened up new possibilities for experimentalists to detect slow moving super heavy MMs via nucleon catalyzing effects, see [7] for a brief review of the detection techniques.

In this short note we would like to draw attention to one peculiar aspect of relativistic MMs detection by deep underwater/ice Cherenkov neutrino telescopes. The detection principle of this kind of MMs is based on the detection of Cherenkov light induced by MMs in water or ice. The number of Cherenkov photons, N, induced by MMs is defined by a well-known formula:

$$dN/d\lambda = (2\pi\alpha/\lambda^2)(ng)^2(1-1/(n^2\beta^2)) \qquad (1)$$

where $\alpha$ is the fine structure constant, $g$ is the magnetic charge, $n$ is the refractive index. The magnetic charge is defined by Dirac's electric charge quantization formula:

$$g = 2\pi\xi\hbar c/e, \ \xi = 0, \pm 1, \pm 2, \qquad (2)$$
$$g = \xi 68.5e \qquad (3)$$

The intensity of Cherenkov radiation due to a relativistic MM with the basic magnetic charge moving in a media with a refractive index $n$ should be $(gn)^2$ times larger than that for a relativistic muon in the same media. The basic strategy of searches for relativistic MMs moving in water or ice boils down to the detection of muon like events with a ~8300 times higher light intensity. It means that one searches for "naked" relativistic MMs. Based on their experimental data the Baikal and AMANDA experiments set rather stringent upper limits on fast MMs with

$0.75 \leq \beta \leq 1$. The upper limits, obtained by both experiments, are shown in Fig.1 [8,9]. The results are undoubtedly very impressive. The AMANDA experimental limit of $6 \cdot 10^{-17}$cm$^{-2}$s$^{-1}$sr$^{-1}$ is actually even lower than the astrophysical limit, the so-called Extended Parker Bound (EPB) of $10^{-16}$cm$^{-2}$sr$^{-1}$s$^{-1}$ [10]. But a closer look to the limits at $\beta = 1$ reveals explicitly the nonphysical character of the limit at this exact point. Of course, it is a matter of a convention accepted by all the persons involved in the field, but on the other hand, when $\beta$ gets closer and closer to unity, one can no longer consider just "naked" MMs. In this case they produce more and more accompanying electromagnetic showers and eventually, at high values of the Lorentz-factor $\gamma$, the electromagnetic showers accompanying such MMs will contribute more to the total Cherenkov light than "naked" MM.

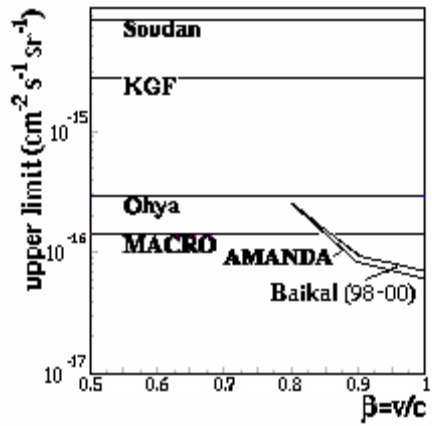

Fig.1. Experimental upper limits on relativistic Magnetic Monopoles, from [9].

It seems that MMs of light or intermediate masses of $10^7$-$10^{10}$ GeV in contrary to GUT MMs with masses of $10^{17}$ GeV or higher, if they exist at all, would be very likely ultra relativistic objects with high values of the Lorentz-factor $\gamma$. Indeed, they would be very easily accelerated by galactic and extragalactic or galactic cluster magnetic fields up to very high energies [11].

$$E_k \sim g \int_{path} B dl \qquad (4)$$
$$E_k \sim g B \xi \eta^{1/2}, \qquad (5)$$

here $E_k$ – MM's kinetic energy; $g$ – the magnetic charge; $B$ – the magnetic field strength; $\xi$ - the magnetic field's coherence length; $\eta$ – the number of coherence field domains along the path. The random walk character is specified by a factor of $\eta^{1/2}$. The value of $\eta$ is roughly estimated to be of the order of 100 [11].

According to (4) and (5) MMs can be accelerated up to the highest energies exceeding even $5 \cdot 10^{23}$ eV [11]. For MMs with masses of $10^7 \div 10^{10}$ GeV the Lorentz-factor can reach $\gamma \geq 10^5 \div 10^8$. MMs with such masses will be ultra relativistic with high values of $\gamma$. For MMs above $\gamma = 10^3$, the energy losses due to direct pair and photonuclear productions in water or ice begin to dominate over other ways of losses, and at $\gamma = 10^6$ the number of charged particles accompanying MMs reaches the value of $\sim 2 \cdot 10^6$ thereby exceeding the direct Cherenkov light intensity due to "naked" MMs by nearly two orders of magnitude. Fig. 2, taken from [11], shows the dependence of the number of accompanying charged particles on the MM's Lorentz-factor. A Lorentz-factor of $10^6$ can be considered as a limit for searches by neutrino telescopes for MMs that are upward going from the bottom hemisphere because likely the Earth is opaque for MMs with a $\gamma$ exceeding $10^6$.

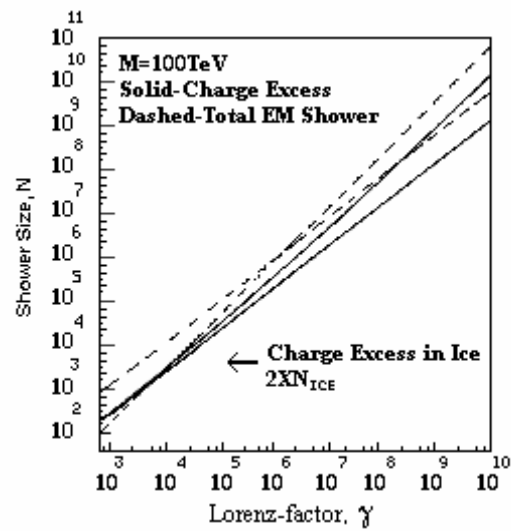

Fig.2. Dependence of shower size on the Lorentz-factor of Magnetic Monopoles [11].

Thus, ultra relativistic MMs with a high value of $\gamma$ should be much brighter objects moving in water or ice and should be seen by optical modules of neutrino telescopes at much larger distances than it was thought so far. In turn the arrays effective areas for MMs detection should be conspicuously larger. As an example we'll try to evaluate upper limits for fluxes of MMs with a $\gamma$ of $10^6$ based on the upper limits obtained by the AMANDA detector for MMs with $\beta = 1$. The AMANDA results are chosen because deep Antarctic ice is much more transparent than deep lake and seawater. The expected effect should therefore be more pronounced. For simplicity only absorption of light is taken into account but no losses due to light scattering, although, of course, the latter losses play an important role in the response of deep underwater/ice neutrino telescopes to MMs. It is necessary to note that we present here just a very naive attempt to demonstrate that real upper limits reached by current underwater/ice experiments might be much more stringent than the limits published at present.

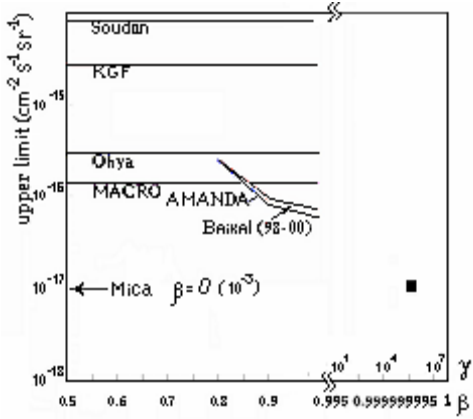

Fig.3. Experimental upper limits on relativistic Magnetic Monopoles, see all references in [9,11] and a hypothetical upper limit re-evaluated here from AMANDA's bound (full rectangle) for a Lorentz-factor $\gamma = 10^6$.

To re-evaluate the present limits on relativistic MMs fluxes one can estimate from which distance MMs, accompanied by intense electromagnetic showers, would be seen by an array's peripheral optical modules (OMs) in comparison with the case of "naked" MMs. The following parameters $R_g$, $R_{nm}$ and $R_\gamma$ are introduced. $R_g$ – a parameter describing a characteristic geometric size of the array. $R_{nm}$ – a characteristic distance at which "naked" MMs are seen by the peripheral OMs of the detector, therefore $(R_g + R_{nm})$ corresponds to an effective area of the detector inferred from the present upper limits on MMs with $\beta = 1$, see Fig.1. $R_\gamma$ is a characteristic distance at which MMs with a large value of $\gamma$ would be seen by the peripheral OMs. $(R_g + R_\gamma)$ corresponds to the new effective area. All these parameters are interconnected by the following trivial expression:

$$R_\gamma exp(R_\gamma/\lambda^*) = (A_\gamma/A_{nm})R_{nm}exp(R_{nm}/\lambda^*), \quad (6)$$

where $A_{nm}$ and $A_\gamma$ are the number of Cherenkov photons for "naked" MMs and MMs with accompanying electromagnetic showers respectively, $\lambda^*$ is absorption length of light. For our estimation we assumed $\lambda^* = 100$ m, $R_g = 60$ m and $R_{nm} = 240$ m.

Therefore, a gain factor might be expressed by the following formula:

$$G=(R_g+R_\gamma)^2/(R_g+R_{nm})^2 \quad (7)$$

According to (6), (7) and Fig.2 for $\gamma = 10^6$ one can get the gain factor $G$ of 5÷6 and a new bound of $\sim \cdot 10^{-17}$ cm$^{-2}\cdot$s$^{-1}\cdot$sr$^{-1}$. It should be noted that this bound is rather close to the bound reached by experiments searching for slow moving MMs with $\beta \sim 10^{-4} \div 10^{-3}$[12] in ancient mica. In Fig.3 the new bound is depicted by a full rectangle for a double scaled $\gamma/\beta$ axis.

In conclusion we would like to emphasize once again the simplification of the approach and on the other hand the absolute necessity to take electromagnetic showers accompanying relativistic MMs into consideration. In this case one should reconsider present limits published already by currently operating Cherenkov experiments for relativistic MMs in order to set more stringent ones. Furthermore it would be very interesting to analyze the capabilities of imaging atmospheric Cherenkov telescopes like MAGIC or HESS to detect ultra relativistic MMs. Their large effective area could very likely overweight their comparatively small duty cycle. MM signatures in such telescopes would be spectacular – very bright muon like circles.

I would like to thank my colleagues and friends from the Baikal and AMANDA experiments for the fruitful discussions and particularly Dr.V.Ch.Lubsandorzhieva and R.V.Vasiliev for many invaluable remarks and help in preparation of this paper.